\pdfoutput=1
\documentclass[sigconf, authorversion=True]{acmart}
\usepackage{multirow}
\usepackage{subcaption}
\usepackage{hyperref}
\AtBeginDocument{%
  }

\copyrightyear{2024}
\acmYear{2024}
\setcopyright{acmlicensed}
\acmConference[RecSys '24]{18th ACM Conference on Recommender Systems}{October 14--18, 2024}{Bari, Italy}
\acmBooktitle{18th ACM Conference on Recommender Systems (RecSys '24), October 14--18, 2024, Bari, Italy}
\acmDOI{10.1145/3640457.3688172}
\acmISBN{979-8-4007-0505-2/24/10}

\makeatletter
\def\@adddotafter#1{%
  \ifx\relax#1\relax
    #1%
  \else
    \def\@tempb{\expandafter\@gobble\string#1.}%
    \ifx\@tempb\@empty
      #1\@addpunct{.}%
    \else
      #1%
    \fi
  \fi
}

\begin{document}

\title{Comparative Analysis of Pretrained Audio Representations in Music Recommender Systems}

% \affiliation{%

\author{Yan-Martin Tamm}
\email{yanmart.tamm@gmail.com}
\orcid{0000-0002-6174-7736}
\affiliation{%
  \institution{University of Tartu}
  \streetaddress{Narva mnt 18}
  \city{Tartu}
  \country{Estonia}
  \postcode{51009}
}

\author{Anna Aljanaki}
\email{aljanaki@gmail.com}
\orcid{0000-0002-7119-8312}
\affiliation{%
  \institution{University of Tartu}
  \streetaddress{Narva mnt 18}
  \city{Tartu}
  \country{Estonia}
  \postcode{51009}
}


% \renewcommand{\shortauthors}{Tamm and Aljanaki}

\begin{abstract}
  Over the years, Music Information Retrieval (MIR) has proposed various models pretrained on large amounts of music data. Transfer learning showcases the proven effectiveness of pretrained backend models with a broad spectrum of downstream tasks, including auto-tagging and genre classification. However, MIR papers generally do not explore the efficiency of pretrained models for Music Recommender Systems (MRS). In addition, the Recommender Systems community tends to favour traditional end-to-end neural network learning over these models. Our research addresses this gap and evaluates the applicability of six pretrained backend models (MusicFM, \mbox{Music2Vec}, MERT, EncodecMAE, Jukebox, and MusiCNN) in the context of MRS. We assess their performance using three recommendation models: K-nearest neighbours (KNN), shallow neural network, and \mbox{BERT4Rec}. Our findings suggest that pretrained audio representations exhibit significant performance variability between traditional MIR tasks and MRS, indicating that valuable aspects of musical information captured by backend models may differ depending on the task. This study establishes a foundation for further exploration of pretrained audio representations to enhance music recommendation systems. 
\end{abstract}




% \received{20 February 20077}

\hyphenation{MusicFM MERT EncodecMAE Jukebox MusiCNN HitRate}

\maketitle
\section{Introduction}

Music Recommender Systems (MRS) are naturally fit for a hybrid recommendation setting because both collaborative interactions and audio data are available, allowing us to gain deeper insight into user preferences. This not only improves performance but also has the potential to address the cold start problem for new items.

In content-aware MRS, Convolutional Neural Networks (CNN) are commonly trained on Mel-Spectrograms extracted from music segments, as popularized by \cite{Oord2013DeepCM}. This approach proved highly effective in processing audio and incorporating content information into recommender systems.

Since then, the Music Information Retrieval (MIR) community has introduced numerous backend models that are pretrained on extensive amounts of music data. Some of these models, such as musiCNN ~\cite{Pons2019MusiCNNPC}, are trained in a supervised manner, usually on auto-tagging, while others, like Jukebox ~\cite{Dhariwal2020JukeboxAG}, are self-supervised. Regardless of the approach, the crucial factor is that these models can effectively be utilized for downstream tasks through transfer learning, yielding results comparable to state-of-the-art models specifically designed for those tasks. This opens up the possibility of using large quantities of unlabeled music data to address problems with limited labeled examples, which could be particularly beneficial in MRS research, where access to large datasets containing both music data and user play history is limited due to copyright.

However, there has been a lack of exploration of pretrained audio representations within the context of MRS. MIR papers typically do not research the effectiveness of pretrained backend models for MRS. At the same time, the Recommender Systems (RS) community tends to lean towards traditional end-to-end neural network learning over these models. One notable exception is \cite{Park2022ExploitingNP}, where the authors utilized three pretrained encoders: CLMR ~\cite{Spijkervet2021ContrastiveLO}, MEE~\cite{Koo2022EndToEndMR}, and Jukebox ~\cite{Dhariwal2020JukeboxAG}. However, the primary focus of that paper was to study the role of negative preferences in user music tastes, and the use of different pretrained models emphasised the stability of the proposed method rather than being integral to the research.

Our paper studies the performance of pretrained embeddings in the context of MRS using six recent pretrained backend models. We outline the following research questions:
\begin{itemize}
    \item RQ1: Are pretrained audio representations a viable option for MRS?
    \item RQ2: How do different backend models compare in the context of MRS?
    \item RQ3: How does pretrained backend model performance in MRS correspond to performance in MIR tasks?
\end{itemize}

The rest of the paper is organized as follows: first, we briefly describe pretrained backend models that will be used to generate audio representations. Then, we describe the dataset and the training details\footnote{The code is available at \href{https://github.com/Darel13712/pretrained-audio-representations}{github.com/Darel13712/pretrained-audio-representations}}. We conclude with results and discussion.

\section{Methods}
\subsection{Pretrained Audio Representations}

The list of the models we used can be found in Table \ref{tab:models}. Further, we describe each of them.

\textfloatsep=2.5ex

\begin{table}
    \centering
\caption{Embedding sizes for pretrained audio representations we used}
\label{tab:models}
    \begin{tabular}{cc}
         Model & Embedding Size \\
 MFCC&104\\
         MusiCNN \cite{Pons2019MusiCNNPC}&  200\\
         MusicFM \cite{Won2023AFM}&  750\\
         EncodecMAE \cite{Pepino2023EnCodecMAELN}&  768\\
         Music2Vec \cite{Li2022MAPMusic2VecAS}&  768\\
         MERT \cite{Li2023MERTAM}&  1024\\
         Jukebox \cite{Dhariwal2020JukeboxAG, Castellon2021CodifiedAL}&  4800\\
    \end{tabular}
\end{table}

Mel Frequency Cepstral Coefficients (MFCCs) represent a short-term power spectrum of a sound based on a linear cosine transform of a log power spectrum on a nonlinear mel scale of frequency. MFCC is a low-level acoustic feature designed to capture the timbral characteristics of an audio signal. It is widely utilized in various domains \cite{Abdul2022MelFC}, including RS. Strictly speaking, MFCC is not a pretrained audio representation since there is no backend model with learnable parameters behind it. However, we employ this precalculated representation as a baseline and a reference point, utilizing the mean and flattened covariance matrix of MFCCs provided with the dataset we have utilized.

MusiCNN \cite{Pons2019MusiCNNPC} is a CNN trained in a supervised way to predict crowd-sourced labels (50 classes: tags from last.fm). It takes log mel spectrograms of audio files as an input and applies a series of convolutional and dense layers to them. It was trained on 200k audio files from the Million Song Dataset ~\cite{BertinMahieux2011TheMS}.

Jukebox \cite{Dhariwal2020JukeboxAG} is a music generation model that consists of three separate Vector Quantized Variational Autoencoders (VQ-VAE) with different temporal resolutions. The encoder part of VQ-VAE compresses raw audio input into a sequence of embeddings using \mbox{1-D convolutions}. This sequence is then turned into a sequence of discrete tokens using codebooks. The decoder reconstructs raw audio from latent representations. Sequences of compressed tokens are then processed by an autoregressive Sparse Transformer to learn the prior to generate further samples. This approach of training a Language Model over tokenized music representation showed to be a robust foundation for downstream MIR tasks \cite{Castellon2021CodifiedAL}.

Music2vec \cite{Li2022MAPMusic2VecAS} uses a multi-layer 1-D CNN feature extractor to compress 16kHz audio input into 50Hz representations that are fed into 12-layer Transformer Blocks. The student model takes partially masked input and tries to predict the average of the top-K layers of the teacher model.

Encodec \cite{Defossez2022HighFN} is a neural audio codec that compresses raw audio from 24kHz to 75Hz. It is done by applying a series of convolutional and LSTM blocks to get a sequence of 128-dimensional vectors, which are processed with a residual vector quantization block (RVQ) that maps the input vector to the index of one of the 1024 closest codebook words, then calculates the residual and maps it to a second codebook and so on for a total of 32 codebooks. EncodecMAE ~\cite{Pepino2023EnCodecMAELN} further adopts these representations and compresses them into a single embedding. That is done by applying a Masked Auto Encoder (MAE) on raw Encodec outputs before the RVQ block to predict discrete targets from the RVQ codebooks.

MERT \cite{Li2023MERTAM} is trained in a masked language modelling paradigm, incorporating teacher models to generate pseudo labels:  an acoustic teacher based on Residual Vector Quantisation — Variational AutoEncoder and a musical teacher based on the Constant-Q Transform. Notably, the model can scale from 95M to 330M parameters.

Musicfm \cite{Won2023AFM} is an improvement over MERT design where a BERT-style encoder~\cite{Hsu2021HuBERTSS} is replaced with a Conformer~\cite{Zhang2020PushingTL} and \mbox{k-means} clustering tokenization is replaced with random projection and random codebook approach from BEST-RQ~\cite{Chiu2022SelfsupervisedLW}.

All models described above produce representations for small chunks of audio with lengths ranging from a couple of milliseconds to a couple of seconds. To get a single track-level representation, we average these embeddings over time.

\subsection{Recommendation models}
After obtaining embeddings for each audio track using pretrained backend models, we must decide how to use them to produce recommendations. It is important to note that the goal of our study is not finding the best possible architecture of a neural network for this task but rather estimating the usefulness of such embeddings using diverse approaches. To this end, we use three methods of increasing complexity:
\begin{itemize}
    \item K-Nearest Neighbours (KNN)
    \item Shallow neural network
    \item BERT4Rec \cite{Sun2019BERT4RecSR}
\end{itemize}

To implement KNN, we create a representation of a user by averaging the embeddings of the items in their profile and then generate recommendations that are close to that average point. While this is a simple approach, it is crucial for our experiment as it allows us to assess the potential amount of valuable information available for a recommendation task in content-based embeddings.

For the next approach, we incorporate the listening history described in section \ref{sec:data}. Specifically, we process user and item IDs with an embedding layer and a fully connected layer preserving dimensions with a ReLU activation. The score for the user-item pair is the cosine between the resulting vectors. The item embedding layer is initialized with pretrained embeddings from the corresponding backend model. Moreover, we freeze the weights for the item embedding layer to preserve useful content information stored in them. The user embedding layer is initialized with a mean of the user's tracks, but the weights are unfrozen and can be changed. The model is trained with Max Margin Hinge Loss as in~\cite{Lee2018DeepCE} and negative sampling strategy from~\cite{Magron2021NeuralCC}. More specifically, for each user-item pair from the dataset, we consider an item and a user as positive examples and sample an additional 20 negative users who did not interact with this item for training. Our preliminary studies tested the configuration for frozen item weights, user initialization, and negative sampling strategy and showed the best results. We refer to this model as Shallow Net.

BERT4Rec \cite{Sun2019BERT4RecSR} is a popular and effective model for sequential recommendations that leverages the Bidirectional Encoder Representations from Transformers (BERT). \mbox{BERT4Rec} learning task is to mask some elements in a sequence and to predict them from the context. We chose it to estimate the potential of a more complex and drastically different architecture in our setting. We incorporate pretrained embeddings as a frozen projection of the embedding module, just as we did with the Shallow Net. Due to memory constraints, we limited the maximum length of a sequence to 300, thus limiting the amount of data used for prediction. This value approximately equals the mean length of a user history in our dataset, so it fully covers more than half of the profiles. However, the results may still be further improved by increasing this parameter to cover the complete profiles of all users. To generate recommendations we ranked all items according to scores predicted at the last timestamp of a user profile.

Our study primarily focuses on comparing pretrained audio representations. As a part of this, we keep the item embeddings frozen, as explained earlier. However, we also present the results with regular item initializations for Shallow Net and \mbox{BERT4Rec} to provide a reference point. This helps us test whether frozen item embeddings enhance the recommendation model or impose an unnecessary constraint that hinders performance. We refer to this as random initialization. In this scenario, item embeddings are randomly initialized, unfrozen, and fully learned using the recommendation model. Another option is to initialize the model with pretrained embeddings and unfreeze the weights, but we have omitted this as the results are similar to the frozen variant.

\subsection{Dataset}
\label{sec:data}
We use the Music4All-Onion \cite{Moscati2022Music4AllOnionA} dataset because it contains both music samples, user listening history and precomputed MFCC features. We select the last month of listening history (everything after 2020-02-20) as test and validation data and the previous year as training data. Test and validation contain only the items that users have not listened to before 2020-02-20, so all the items in the test and validation set are new in a user's listening history. We split last month into validation and test set by users in half. Further, we ensure that validation and test data do not contain new users and items that do not appear in train data. Initially, we also planned to test the performance on cold items. However, the particular properties of this dataset are not well suited for this since there are not enough cold items in the last month for proper evaluation. Resulting split sizes can be found in Table \ref{tab:data}.

\begin{table}
    \centering
\caption{Train-test split for Music4All-Onion dataset. We used last month for validation and testing, splitting users equally into both groups at random. The previous 12 months are used as train data. Cold users and items are removed.}
\label{tab:data}
    \begin{tabular}{cccc}
         &  Train&  Validation& Test\\
         Num Users&  17,053&  6,092& 6,092\\
         Num Items&  56,193&  36,942& 37,797\\
 Num Interactions& 5,122,221& 132,425&138,299\\
    \end{tabular}

\end{table}

\subsection{Training Details}
We use Adam optimizer with lr=0.001, early stopping, and we reduce the learning rate on the plateau. For Shallow Net, we train for 100 epochs with cosine-based Hinge Loss. For each positive user-item pair, we sampled 20 negative users at random for this item. For \mbox{BERT4Rec}, we train for 200 epochs with Cross Entropy loss. We calculate HitRate@50, Recall@50 and NDCG@50 to evaluate the results. We also computed MRR and Precision, but we omit them from this paper because they are highly correlated with the other metrics and do not give a broader perspective. We remove listened tracks from recommendations with each model to predict only new items.

\section{Results}
Table \ref{tab:performance} shows the performance of audio representations using different recommendation models.

\begin{table}[htbp]
    \centering
    \caption{Comparison of performance of different pretrained embeddings using KNN, Shallow Net, and \mbox{BERT4Rec} models to generate recommendations. Random value in the embeddings column corresponds to random initialization of item embeddings in a model. With random initialization, item embeddings are unfrozen and can be learned like usual.}
    \begin{tabular}{lcccccc}
    \toprule
    Embeddings & HitRate@50 & Recall@50 & NDCG@50 \\
    \midrule
    \multicolumn{4}{c}{KNN} \\
    \midrule
    MusicFM & 0.009 & 0.000 & 0.000 \\
    MFCC & 0.028 &  0.001 & 0.001 \\
    Music2Vec & 0.033  & 0.002 & 0.001 \\
    MERT & 0.049 &  0.003 & 0.002 \\
    EncodecMAE & 0.054 & 0.003 & 0.002 \\
    Jukebox & 0.057  & 0.003 & 0.002 \\
    MusiCNN & \textbf{0.089} & \textbf{0.005} & \textbf{0.004} \\
    \midrule
    \multicolumn{4}{c}{Shallow Net} \\
    \midrule
    Random & 0.021 & 0.001 & 0.001 \\ 
    MusicFM & 0.108 & 0.007 & 0.005 \\
    MFCC & 0.226 & 0.018 & 0.013 \\
    Music2Vec & 0.291 & 0.029 & 0.021 \\
    MERT & 0.291 & 0.030 & 0.021 \\
    EncodecMAE & 0.296 & 0.031 & 0.021 \\
    Jukebox & 0.272  & 0.029 & 0.020 \\
    MusiCNN & \textbf{0.329} & \textbf{0.037} & \textbf{0.025} \\
    \midrule
    \multicolumn{4}{c}{BERT4Rec} \\
    \midrule
    Random & 0.348 & 0.049 & 0.038 \\
    MusicFM & 0.261  & 0.021 & 0.016 \\
    MFCC & 0.231  & 0.019 & 0.014 \\
    Music2Vec & 0.281 & 0.025 & 0.020 \\
    MERT & 0.360  & 0.051 & 0.038 \\
    EncodecMAE & 0.349  & 0.050 & 0.038 \\
    Jukebox & 0.219  & 0.015 & 0.012 \\
    MusiCNN & \textbf{0.385}  & \textbf{0.058} & \textbf{0.044} \\
    \bottomrule
    \end{tabular}
    \label{tab:performance}
\end{table}

The first observation is that, on average, more complicated recommendation models tend to give better performance. Across different pretrained representations, Shallow Net performs better than KNN, and \mbox{BERT4Rec} performs better than Shallow Net, which is expected but highlights the importance of choosing a model architecture with appropriate complexity.

Secondly, combining content and collaborative information tends to improve results over pure collaborative variant. This is true for all backend models used with Shallow Net. The poor performance of Shallow Net with random initialization can be seen as a disadvantage. However, in our experiment, it is rather an advantage since all the boost in performance with content embeddings comes from the information stored in them. The fact that all backend models enriched with collaborative information show~10 times better performance than their respective raw KNN variants, shows the synergy between content embeddings and collaborative data.

However, with \mbox{BERT4Rec}, results vary for different embeddings. The model by itself shows good performance. MusiCNN shows a statistically significant (p < 0.05) improvement over the base \mbox{BERT4Rec}. MERT and EncodecMAE perform comparably to collaborative \mbox{BERT4Rec} with random initialization, but the difference is not statistically significant. The performance of MusicFM, \mbox{Music2Vec} and Jukebox is worse than that of the pure collaborative variant. This indicates that it is harder for \mbox{BERT4Rec} to extract useful information from them, and a more elaborate procedure of inferring content knowledge should be employed.

If we compare the performance of different backend models across all the approaches, we can see that MusicFM tends to end up in lower positions, even lower than MFCC for KNN and Shallow Net. Music2Vec tends to be slightly better. Jukebox is the second best option with KNN and Shallow Net but significantly drops with \mbox{BERT4Rec}, probably because it has a dimension size 4800, which is much bigger than other models. MERT and EncodecMAE show similar performance overall, but MERT works better with \mbox{BERT4Rec}. MusiCNN constantly shows the best performance across all tests.

\subsection{Comparison to MIR results}

\begin{table}
    \centering
\caption{Comparison of backend models applied to different MIR tasks and MRS. Results for MIR are taken from the respective papers of each model; the last column is from this paper.}
\label{tab:mir}
    \begin{tabular}{ccccc} 
         Model&  Tags&  Genres&  Key& Recs\\ 
         MusicFM&  \textbf{0.924}&  —&  \textbf{0.674}& 0.261\\ 
         Music2Vec&  0.895&  0.766&  0.508& 0.281\\ 
         MERT&  0.913&  0.793&  0.656& 0.360\\ 
         EncodecMAE&  —&  \textbf{0.862}&  —& 0.349\\ 
         Jukebox&  0.915&  0.797&  0.667& 0.219\\ 
         MusiCNN& 0.906& 0.790& 0.128&\textbf{0.385}
    \end{tabular}
\end{table}
We used self-reported data from corresponding papers to compare our results against the performance of backend models on MIR tasks. Specifically, in Table \ref{tab:mir} Tags column is the AUC metric on MagnaTagATune~\cite{Law200910TI}, Genres is the genre classification accuracy on GTZAN~\cite{Tzanetakis2002MusicalGC}, Key is the key detection accuracy on Giantsteps~\cite{Knees2015TwoDS}, and Recs is the recommendation HitRate@50 with \mbox{BERT4Rec} reported in this paper. We tried to choose the tasks and datasets reported by all the models we used. However, EncodecMAE results are unavailable for tag prediction and key estimation, and MusicFM results are unavailable for genre prediction.

We can see a drastic difference when we compare the performance of backend models in MIR tasks and MRS. MusicFM and Jukebox hold the best results in auto-tagging and key prediction but are the worst for recommendations. However, MERT, the third-ranking model for other MIR tasks, is the second-best model for recommendations, which suggests that it contains valuable information for both tasks. The same goes for EncodecMAE, which shows the best performance in genre prediction and the third best in recommendations. \mbox{Music2Vec} tends to show worse results across all tasks, both MIR and MRS. A surprising difference can be found with MusiCNN because it is comparable to MusicFM and Jukebox but has slightly lower results for tags and genres, the lowest performance for key detection, and the best performance for recommendations across all our experiments. The low performance of MusicFM in our evaluations and high performance in the MIR context might suggest technical problems with published weights for the model we used since MusicFM is a modification over MERT, which shows good performance across all tasks.

\subsection{Discussion}

\subsubsection{RQ1: Are pretrained audio representations a viable option for MRS?}

Our experiments show that improving a pure collaborative model with content information is possible without model finetuning or end-to-end re-learning, thus advocating for broader usage of pretrained backend models in MRS. 

\subsubsection{RQ2: How do different backend models compare in the context of MRS?}

MusiCNN shows consistently good results in all our tests, which suggests that the supervised auto-tagging task it was trained on contains a lot of useful information for MRS. Two other good options for MRS are MERT and EncodecMAE. \mbox{Music2Vec} shows slightly worse performance, which aligns with its performance in other MIR tasks. MusicFM shows inferior performance in our tests but outstanding performance in MIR, which may suggest some technical problems with the published model weights that we used. Jukebox's performance is better than most models with KNN but tends to fall in position relative to other models with the increasing model complexity in our experiments. Combined with good results in MIR, it may suggest a need for more elaborate embedding processing than we used, possibly because the embedding size is much larger.

\subsubsection{RQ3: How does pretrained backend model performance in MRS correspond to performance in MIR tasks?}

The performance can vary between different downstream tasks, the most notable difference being that MusiCNN is showing outstanding results in MRS.

\subsection{Limitations}
One of the limitations of our work is the usage of only one dataset, which may undermine the generalizability of our results. The second limitation is the number and scope of the recommendation models studied. Our results show that model architecture plays a significant role in the amount of useful information extracted from content embeddings, which calls for a broader scope of models. Moreover, we studied only one way to incorporate embeddings into a recommender system by using them as frozen item embeddings with a learned transformation over them. However, it is also possible to try other approaches, like predicting collaborative embeddings using content information~\cite{Oord2013DeepCM} or using content embedding as a regularization on collaborative one~\cite{Magron2021NeuralCC}. Our work can be further improved by comparing end-to-end CNN models widespread in MRS with general pretrained MIR models. Finally, a notable direction for further work is measuring the performance of our approach in the cold-start scenario.

\section{Conclusion}

We compared different frozen backend models for an MRS task using three ways to incorporate them into the recommendation process. We showed it is a viable approach to improve pure collaborative model performance. We found that EncodecMAE, MERT and MusiCNN performed well in the context of MRS. Comparing the performance of these models in MRS and MIR tasks, we demonstrate that best-performing MIR models do not always translate to best-performing MRS models. Notably, the supervised tag prediction task of MusiCNN suggests the usefulness of tags like genres, instruments, and emotions in improving recommendations. We hope this paper proves helpful in inspiring the adoption of pretrained audio representations in MRS.

\bibliographystyle{ACM-Reference-Format}
\bibliography{bibliography}

\end{document}